\documentclass[12pt]{article}
\usepackage{amsmath,amsthm,amsfonts,braket,algorithm,graphicx}
\usepackage{fullpage,authblk}

\theoremstyle{definition} 
\theoremstyle{definition} 
\newtheorem {theorem} {Theorem}

\newcommand{\kb}[1]{\mathbf{[#1]}}

\newcommand{\bk}[1]{\braket{#1|#1}}

\newcommand{\given}{\text{ }|\text{ }}

\newcommand{\M}{\texttt{Measure}}
\newcommand{\R}{\texttt{Reflect}}

\newcommand{\rej}{\sigma_{\text{reject}}}

\title{Improved Semi-Quantum Key Distribution with Two Almost-Classical Users}

\date{}
\author[1]{Saachi Mutreja}
\author[2]{Walter O. Krawec\footnote{Email: \texttt{walter.krawec@uconn.edu}}}
\affil[1]{\small{UC Berkeley}\\\small{Berkeley, CA 94720, USA}}
\affil[2]{\small{Department of Computer Science and Engineering}\\\small{University of Connecticut}\\\small{Storrs, CT 06269 USA}}

\begin{document}

\maketitle

\begin{abstract}
Semi-quantum key distribution (SQKD) protocols attempt to establish a shared secret key between users, secure against computationally unbounded adversaries.  Unlike standard quantum key distribution protocols, SQKD protocols contain at least one user who is limited in their quantum abilities and is almost ``classical'' in nature.  In this paper, we revisit a mediated semi-quantum key distribution protocol, introduced by Massa et al., in 2019, where users need only the ability to detect a qubit, or reflect a qubit; they do not need to perform any other basis measurement; nor do they need to prepare quantum signals.  Users require the services of a quantum server which may be controlled by the adversary.  In this paper, we show how this protocol may be extended to improve its efficiency and also its noise tolerance.  We discuss an extension which allows more communication rounds to be directly usable; we analyze the key-rate of this extension in the asymptotic scenario for a particular class of attacks and compare with prior work.  Finally, we evaluate the protocol's performance in a variety of lossy and noisy channels.
\end{abstract}

\section{Introduction}

Semi-quantum key distribution (SQKD) allows two parties, Alice and Bob, to establish a shared secret key that is secure against computationally unbounded adversaries.  Unlike standard quantum key distribution, with SQKD, at least one party is restricted to being ``classical'' in nature - namely, at least one party is restricted to operating in a single, publicly known, basis, or to disconnecting from the quantum channel and reflecting all qubits back to the sender.  This limited ``classical'' party is not permitted to measure or send qubits in arbitrary bases.  Semi-quantum key distribution was introduced originally in 2007 in \cite{SQKD-first} and, since then, has led to a growing area of research interest with new protocols and security proofs for both QKD \cite{SQKD-second,zou2009semiquantum,mdi-1,amer2019semiquantum,vlachou2018quantum,iqbal2020high,boyer2017experimentally,krawec2018practical,silva2021semi,chongqiang2022efficient} and alternative cryptographic primitives such as secret sharing \cite{secret-1,secret-2,secret-3}, secure direct communication \cite{SDC-1,SDC-2,SDC-3,SDC-4,SDC-5}, private  comparison \cite{priv-comp-1,priv-comp-2}, and secure identification protocols \cite{ident-1,ident-2}.  It is also experimentally realizable \cite{massa2019experimental,gurevich2013experimental}.  For a general survey of semi-quantum cryptography, the reader is referred to \cite{SQKD-survey} while for a general survey of QKD, the reader is referred to \cite{qkd-survey-pirandola,qkd-survey-scarani,amer2021introduction}.

Recently, a form of \emph{mediated semi-quantum key distribution} (M-SQKD) protocol (a model originally introduced in \cite{krawec2015mediated}) was developed in \cite{massa2019experimental}.  Here, two parties wish to establish a shared secret key; however these parties are only able to detect the presence of a photon, or to reflect a photon back to a sender.  They cannot even prepare new quantum signals.  Clearly, such a protocol cannot be secure (or even correct) without the help of a third-party who is capable of performing some alternative quantum operations.  This third-party, called a quantum server, is responsible for creating the initial quantum state, and later performing a particular quantum measurement and reporting the outcome.  This protocol was also experimentally implemented.  Interestingly, as proven in \cite{massa2019experimental}, the server need not be trusted and may in fact be adversarial; security is still possible even though users are so restricted.  A variant of this protocol was shown to be partially device independent in \cite{silva2021semi}.

The protocol described in \cite{massa2019experimental} (which we call here MZ-M-SQKD19 as it is a Mach-Zehnder based Mediated SQKD protocol developed in 2019) though proven secure in the finite-key setting under practical device constraints (e.g., imperfect detectors and weak coherent sources) and collective attacks, was inefficient.  Under ideal scenarios, the key-rate of the protocol was only $1/8$ meaning that $8$ qubits were required to distill $1$ secret key bit assuming no noise or loss (if there is noise and/or loss, the key-rate, of course, drops even more).  This inefficiency is due to the fact that users must discard numerous rounds and only use particular rounds where things ``go right'' (namely, a random measurement from the server produces the right outcome).

In this work,  we revisit this original protocol of \cite{massa2019experimental} and extend it to increase its efficiency.  We also demonstrate that our extension can increase the protocol's noise tolerance.  Our work is primarily concerned with improving the efficiency of this original protocol and, to evaluate, we conduct an information theoretic proof of security assuming single photons and lossy channels, though assuming an adversarial server.  We compute the protocol's key-rate in the asymptotic scenario under a particular form of i.i.d. attack.  Our extension adds a potential second ``sub-round'' for every protocol round; this greatly complicates the security analysis and our methods may be useful in other (S)QKD protocols.  Importantly, our methods in improving the efficiency of this M-SQKD protocol may be useful to other experimentally realizable semi-quantum protocols, such as \cite{boyer2017experimentally,krawec2018practical} as we show how previously discarded events may be utilized in the semi-quantum model, through careful use of a second sub-round.  Finally, we evaluate the protocol's performance in a variety of scenarios including noisy and lossy channels and adversarial servers.

\section{Notation and Preliminaries}

Given a quantum state $\rho_A$ (a Hermitian positive semi-definite operator of unit trace) acting on some Hilbert space $\mathcal{H}_A$.  We write $H(A)_\rho$ to mean the von Neumann entropy of the system.  Namely $H(A)_\rho = -tr(\rho_A\log\rho_A)$, where all logarithms in this paper are base two unless otherwise specified.  Given a bipartite state $\rho_{AE}$, we write $H(A|E)_\rho$ to be the conditional von Neumann entropy, namely $H(A|E)_\rho = H(AE)_\rho - H(E)_\rho$.  If both systems are classical, then $H(A|E)_\rho$ is the Shannon entropy (in which case we will often forgo writing the subscript when there is no ambiguity).  We use $h(x)$ to mean the binary Shannon entropy, namely $h(x) = -x\log x - (1-x)\log(1-x)$. Given a pure state $\ket{\psi}$, we write $\kb{\psi}$ to mean $\ket{\psi}\bra{\psi}$. 

In general, QKD protocols (semi-quantum or otherwise), will first utilize the quantum channel and authenticated classical channel to establish a \emph{raw key} of size $N$ bits; this process requires sending $M \ge N$ qubits (in general, $N = p_{acc}M$, where $p_{acc}$ is the probability that a round is ``accepted'' and not discarded by users).  These raw keys are classical bit strings, one held by Alice and another by Bob, which are partially correlated and partially secret.  Following this, an error correction protocol and privacy amplification protocol are run, establishing a final secret key of size $\ell$ bits.  Two important metrics for any QKD protocol are its \emph{key rate} $r = \ell/N$ and its \emph{effective key rate} $r' = \ell/M$.  In the asymptotic scenario, where $M \rightarrow \infty$, and assuming collective attacks, we may use the Devetak-Winter keyrate equation \cite{QKD-Winter-Keyrate,QKD-renner-keyrate} to evaluate these rates leading to:
\begin{equation}\label{eq:key-rate}
  r = H(A|E)_\rho - H(A|B)
\end{equation}
The effective key-rate, $r'$, is typically found by relating the number of qubits sent to the size of the raw key (e.g., if $N = p_{acc}M$, then $r' = p_{acc}r$).  Above, $\rho_{ABE}$ is the state of the system modeling a single quantum communication round, conditioned on it being accepted (i.e., conditioned on it leading to a raw key bit being generated).

To actually compute the key-rate, we will therefore need a bound on the entropy $H(A|E)_\rho$ and $H(A|B)$.  The latter is typically easy to compute directly since it is a function of Alice and Bob only (who, through standard sampling arguments, may fully know their joint $AB$ distribution and thus evaluate $H(A|B)$ directly).  Bounding the quantum entropy $H(A|E)_\rho$ is the more difficult challenge and usually the key ingredient in any security proof.  For that, we will later use the following result from \cite{QKD-Tom-Krawec-Arbitrary}:

\begin{theorem}\label{thm:entropy}
  (From \cite{QKD-Tom-Krawec-Arbitrary} though generalized for our application here): Given a quantum state $\rho_{AE}$ of the form:
  \begin{equation}
    \rho_{AE} = \frac{1}{N}\kb{0}_A \otimes \left(\sum_{i=0}^m \kb{E_i}\right) + \frac{1}{N}\kb{1}_A \otimes \left(\sum_{i=0}^m\kb{F_i}\right),
  \end{equation}
  then for every $0 \le m' \le m$ it holds that:
  \begin{equation}
    H(A|E)_\rho \ge \sum_{i=0}^{m'}\left(\frac{\bk{E_i} + \bk{F_i}}{N}\right)\cdot\left(h\left(\frac{\bk{E_i}}{\bk{E_i} + \bk{F_i}}\right) - h(\lambda_i)\right),
  \end{equation}
  with:
  \begin{equation}
    \lambda_i = \frac{1}{2}\left(1 + \frac{\sqrt{(\bk{E_i} - \bk{F_i})^2 + 4|\bk{E_i|F_i}|^2}}{\bk{E_i}+\bk{F_i}}\right).
  \end{equation}
\end{theorem}

Note that the above is a slight generalization of the Theorem as presented in \cite{QKD-Tom-Krawec-Arbitrary}; for a proof that this indeed follows, the reader is referred to \cite{krawec2019multi}.

\section{The Protocol}

\begin{figure}
    \centering
    \includegraphics[width=.5\textwidth]{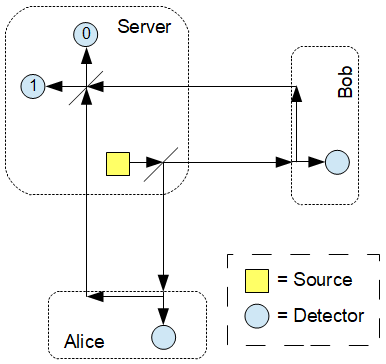}
    \caption{High-level diagram of the protocol assuming an honest server. The server should send a single photon through a beamsplitter causing the qubit to travel towards Alice and Bob in a superposition.  Alice and Bob may independently choose to $\R$ or $\M$.  If $\R$, the signal is sent directly back to the server; otherwise, if $\M$, the signal is routed to a photon detector.  Note that Alice and Bob cannot create new quantum systems.  The server, on return, should pass the signal through a second beamsplitter and report which of the two detectors (if any) received a signal.  The key is derived from users choices - in particular, when users choose \emph{opposite} actions.}
    \label{fig:protocol}
\end{figure}

We now describe the protocol in detail.  We assume that a two way quantum channel connects the server ($C$) to Alice and Bob.  In the ideal scenario, this should constitute a folded Mach-Zehnder interferometer, however since the server may be adversarial, we do not assume this in our security proof later.  An authenticated classical channel connects Alice and Bob; an unauthenticated classical channel connects the server to Alice and Bob. See Figure \ref{fig:protocol}.  We will call our protocol here MZ-M-SQKD19-ext (to distinguish from the original one in \cite{massa2019experimental} which we extend and which, as mentioned, we refer to as MZ-M-SQKD19).

We introduce the protocol assuming an honest server for clarity, however will later prove security against a potentially corrupt server.  The protocol consists of $N$ independent \emph{rounds}, each round consists of two \emph{sub-rounds} with the second sub-round only being used in certain circumstances.  The original protocol from \cite{massa2019experimental} consists only of the first sub-round; our extension here adds this second sub-round in an effort to improve efficiency.  A single round consists of the following process:
\begin{enumerate}
  \item The server $C$ prepares a quantum state of the form $\frac{1}{\sqrt{2}}(\ket{0} + \ket{1})$, with $\ket{0}$ representing a photon traveling to Alice and $\ket{1}$ representing a photon traveling to Bob.  Such a state may be created by sending a single photon through a beamsplitter in a Mach-Zehnder interferometer.  See Figure \ref{fig:protocol}.
  \item Alice and Bob, independently, choose to $\R$ or $\M$.  If $\M$, the photon is routed to a photon detector; otherwise, it is routed back to the server.  If a party chooses $\M$ and detects the photon, they will later signal to discard this round.  Otherwise, if the measuring party does not detect the photon, Alice will use the raw-key encoding scheme that a choice of $\R$ means a key-bit of $0$ and $\M$ means a key-bit of $1$; Bob will use the opposite encoding scheme.  The goal of the protocol, at this point, is for Alice and Bob to guess at the action of the other party without the server (or another third-party adversary) determining what choice was actually made.  Notably, it is the actions of the parties that dictate their raw key, not an actual measurement outcome.
  \item The server will pass the returning signal through the second half of the interferometer (again, see Figure \ref{fig:protocol}) and report the measurement outcome, either ``$0$,'' ``$1$,'' or ``$vac$.''  Here the message ``$vac$'' is used to indicate that the server did not detect a photon (i.e., the server detected the vacuum state).  Normally, if both parties choose $\R$, the interferometer should be calibrated so that the message ``$0$'' is always sent.  Of course natural noise (or adversarial interference) will alter this and any other value will be considered noise that must be taken into account when deriving the key rate.
  \item If the server sends the message ``$vac$'', then Alice and Bob discard this round and repeat from step 1 with a new round; if the server sends the message ``$1$'' then Alice and Bob use this round to contribute towards their raw key and users are finished with this round, proceeding to the next (starting above at step 1).  Ideally, if the server sends the message ``$1$,'' users can be certain they chose \emph{opposite} actions.  See Table \ref{tbl:sub-round1}.
  \item \textbf{Extension: }Otherwise, if the server sends the message ``$0$'', then parties will begin \emph{Sub-Round 2}.  The server will repeat the above process (from step 1) but Alice and Bob will invert their action choice.  In this sub-round, when the server sends its second message, they will reject this entire round only if the server sends ``$vac$'' - otherwise, if the server sends the message ``$0$'' or ``$1$'', they will use this round to contribute towards their raw key.  In particular, they will use the encoding chosen in sub-round 1.
\end{enumerate}
Note that, importantly, Alice and Bob's choice of actions are independently chosen each round; however if the second sub-round is used, their actions depend on their choice in the first sub-round.

\begin{table}[]
\begin{tabular}{llll|l}
$A$ & $B$ & $A_{key}$ & $B_{key}$ & $C$'s Message \\\hline
$\R$ & $\M$ & $0$ & $0$ & ``$0$'', ``$1$'', or ``$vac$''  \\
$\M$ & $\R$ & $1$ & $1$ & ``$0$'', ``$1$'', or ``$vac$''  \\
$\R$ & $\R$ & $0$ & $1$ & ``$0$''  \\
$\M$ & $\M$ & $1$ & $0$ &  ``$vac$'' 
\end{tabular}
\caption{Showing the possible outcomes of a single sub-round of the protocol under analysis assuming ideal conditions and an honest server.  This table applies to both the original MZ-M-SQKD19 and our extension here.  Alice and Bob's keys are derived from their actions.  Note that, whenever the server sends the message ``$1$,'' users can be certain they chose \emph{opposite} actions, thus the need to reverse the action-to-key encoding for Alice and Bob.  In the original protocol, any other message from the server was discarded as it was inconclusive for users.  Here, we propose an extension so that whenever the server sends the message ``$0$,'' users run a second sub-round where they flip their actions.  This second sub-round is discarded only if the server sends the message ``$vac$'' or users detect the photon in a measurement.  See text for greater explanation.}\label{tbl:sub-round1}
\end{table}

After the completion of a round (which may consist of both sub-rounds or just the first sub-round depending on the server's message), the protocol repeats with a new round.  Following the completion of a sufficient number of rounds (in this paper we will consider the asymptotic scenario where the number of rounds goes to infinity), Alice and Bob will disclose a random subset of all measurement outcomes and choices to perform parameter estimation.  In particular, users will choose a random subset and disclose all actions and results on those rounds chosen. Any key material from those rounds chosen for parameter estimation are, of course, discarded.  Following this, assuming the error rate is ``low enough'' (to be discussed), they will perform an error correction and privacy amplification process to distill their final secret key.

We comment that this protocol extends the original MZ-M-SQKD19 protocol from \cite{massa2019experimental} by adding the additional sub-round 2.  That is, the original protocol consisted of steps 1-4 above; our extension adds the additional step 5, repeating the above for a second sub-round  The original protocol would reject any round where the server did not send the message ``$1$.''  Our protocol extension here, by utilizing a second sub-round where users flip their action choice, allows for the potential contribution of message ``$0$'' to be used towards the raw key.  As we show later, this extension can greatly improve the secret key generation rate of this protocol, even when counting for the potential need to send two photons on a single round (i.e., even the effective key rate is improved with our extension).  Interestingly, our extension also improves the noise tolerance of the protocol as we show later.

It is clear that our protocol is correct - namely, in the absence of noise and if the server is honest, then Alice and Bob will agree on the same raw-key.  Indeed, under ideal conditions, the only time the server can send the message ``$1$'' is if one of Alice or Bob, but not both, chose $\M$ and the measuring party did not detect the photon.  Furthermore, in this event, it is always clear to users that the other party choose the opposite action.  Now, in the event the server sends the message ``$0$'' on sub-round 1, it is not immediately clear to parties whether they choose opposite actions or if both parties choose $\R$ (see Table \ref{tbl:sub-round1}).  Thus, the original protocol \cite{massa2019experimental} discarded this event leading to waste.  Our extension, by utilizing a second sub-round \emph{where users flip their actions} can potentially salvage these rounds by having the server send a second qubit and parties flipping their action choices.  Note that, the ambiguity of a message ``$0$'' arises only if both parties choose $\R$ (in which case the server will always send ``$0$'' in ideal conditions).  Thus, by flipping their actions in this case, both parties, in sub-round $2$ will choose $\M$ causing the server to always send the message ``$vac$'' (ideally) in which case parties discard the round.  However, if one party chose $\M$ and the other chose $\R$, in the next sub-round they will choose $\R$ and $\M$ respectively; thus any message of ``$0$'' or ``$1$'' by the server in this second sub-round lets parties know they are choosing opposite actions without ambiguity, thus leading to a correlation for their key.

The reader may now wonder if this is really more efficient than the original protocol in terms of number of photons sent since, for some rounds, two photons are required.  We show later that our protocol, even with adversarial noise, can be more efficient than the original.

\section{Security Analysis}

The goal of this section is to compute a bound on the key-rate of our protocol.  From Equation \ref{eq:key-rate}, this involves, primarily, computing a bound on the von Neumann entropy of the system.  To do this, we must first model a single round of the protocol, conditioning on a key-generation event (from which $H(A|E)$ must be computed).  In our security analysis, we will assume single qubits and lossy channels - that is, we do not consider multi-photon events.  These are important to consider, of course, and though considered for the original protocol \cite{massa2019experimental}, we only consider single qubits and lossy channels here in order to demonstrate how improvements may be made to the protocol in theory, leaving practical issues as interesting future work.  We will also consider only i.i.d. attacks on each sub-round.  In particular, each sub-round will consist of the same (potentially probabilistic) attack operation.  In general, this is weaker than a collective attack which would have the second sub-round attack dependent on the first.  However, we feel that analyzing security even in this case is still a notable contribution and furthermore, if one were to consider the multi-mediated SQKD model introduced in \cite{krawec2019multi}, or a variant of the protocol where users ``shuffle'' individual rounds into appropriate sub-rounds when needed, then it is equivalent to a general collective attack.  Analyzing full collective attacks for a single server without this shuffling process would be interesting future work (though out of scope for this paper), and our method below may serve as a foundation for such an analysis.  Finally, we note that any third party adversary attack may be ``absorbed'' into an adversarial server's attack.  Thus, in our security analysis, we only consider adversarial servers - any third party adversary's attack will be analyzed also as a consequence.

Based on these assumptions, the server will begin the protocol by sending a state of the form:
\[
\ket{\phi_0} = \alpha\ket{0, c_0} + \beta\ket{1,c_1} + \gamma\ket{v, c_v}
\]
Note that the $\alpha, \beta,$ and $\gamma$ may be real numbers as any alternative phase may simply be absorbed into the corresponding $\ket{c_i}$ ancilla state.  Above, $\ket{0}$ represents a single photon traveling towards Alice; $\ket{1}$ represents a photon traveling towards Bob; and $\ket{v}$ represents a vacuum state.  The $\ket{c_i}$ states are arbitrary ancilla states that the (adversarial) server may use to attempt to extract information later.   If Alice chooses to $\M$, then, with probability $\alpha^2$, Alice observes the photon and, ultimately, the round will be discarded.  Otherwise, with probability $1-\alpha^2$, Alice does not observe the photon in which case it collapses to $(\beta\ket{1,c_1} + \gamma\ket{v,c_v})/\sqrt{1-\alpha^2}$.  This will then be the state that returns to the server.  If Bob chooses $\M$, similar identities may be derived.  Of course if both parties choose $\M$ and neither detect the photon, then it collapses to $\ket{v,c_v}$, an event that happens with probability $\gamma^2$.

Following Alice and Bob's actions, a quantum signal, or a vacuum state, returns to Eve.  From this, she may apply any quantum operation; however, she must send a classical message to Alice and Bob.  We may assume that this message is the same to both parties (that is, Eve does not send one message to Alice and a different message to Bob on a single round) - this is easily enforced by having Alice forward all messages she receives from the server directly to Bob over the authenticated channel and any discrepancies will cause Bob to signal to abort the protocol.

As shown in \cite{krawec2015mediated,massa2019experimental}, the most general way to model such an attack is through a quantum instrument \cite{davies1970operational} which, using standard techniques \cite{wilde2011classical}, may be dilated to a unitary operator.  This attack, as shown in \cite{krawec2015mediated}, then consists of Eve taking the return signal, applying an isometry $U$ mapping it to a state living in some Hilbert space $\mathcal{H}_{cl}\otimes\mathcal{H}_E$, where in this case $\mathcal{H}_{cl}$ is spanned by $\{\ket{0}, \ket{1}, \ket{v}\}$ where these three basis states represent the three possible messages Eve could send to Alice and Bob.  Following the application of $U$ to the returned signal, Eve measures the $cl$ register - the outcome determines the message she sends to the parties while the post measured $E$ portion represents her ancilla in this event.  For a proof that this is identical to a general quantum instrument attack, see \cite{krawec2015mediated}.  We describe the action of this attack operator using the following notation:
\begin{align*}
U\ket{0,c_0} &= \ket{0}_{cl}\ket{e_0}_E + \ket{1}_{cl}\ket{e_1}_E + \ket{v}_{cl}\ket{e_v}_E\\
U\ket{1,c_1} &= \ket{0}_{cl}\ket{f_0}_E + \ket{1}_{cl}\ket{f_1}_E + \ket{v}_{cl}\ket{f_v}_E\\
U\ket{v,c_v} &= \ket{0}_{cl}\ket{g_0}_E + \ket{1}_{cl}\ket{g_1}_E + \ket{v}_{cl}\ket{g_v}_E\\
\end{align*}
Note that, in the following text, we often forgo writing the subscripts in the above states.

There are four main paths which can lead to a key being distilled based on the choices of Alice and Bob.  Consider, first, the case when Alice and Bob both choose $\R$ (in which case, should the round be accepted and not rejected, Alice will have a key-bit of $0$ and Bob a key-bit of $1$ - note this is an error event and so, ideally, the probability of it being discarded should be one or close to one).  We trace the protocol's execution in this event in order to derive a density operator describing the state of Alice, Bob, and Eve's registers in this case along with any public communication sent.  Of course, as we only care about cases that lead to a key bit being distilled, we condition on events leading to a successful key generation event.  In the first sub-round, since both parties choose $\R$, the state returns to the server unchanged, namely the state returning is $\ket{\psi_0}$.  Note that, as is normal in QKD security proofs, we assume all noise is caused by the adversary's attack and that, in fact, the adversary replaces the noisy quantum channel with a perfect channel, allowing her to ``hide'' within the natural noise.  Thus, in the event both parties choose $\R$, the state returning is unchanged.  Eve at this point applies $U$ evolving the state to:
\begin{align*}
U\ket{\psi_0} &= \ket{0}_{cl}(\alpha\ket{e_0} + \beta\ket{f_0} + \gamma\ket{g_0})\\
&+\ket{1}_{cl}(\alpha\ket{e_1} + \beta\ket{f_1} + \gamma\ket{g_1})\\
&+\ket{v}_{cl}(\alpha\ket{e_v} + \beta\ket{f_v} + \gamma\ket{g_v}).
\end{align*}
Now, the protocol discards the round if the server sends the message ``$v$'' while, if the server sends the message ``$1$'' they will use this round immediately and proceed to the next round.  In this case, the state of the system is:
\[
\kb{01}_{AB} \otimes \left(\kb{1}_{cl}\otimes P\left(\alpha\ket{e_1} + \beta\ket{f_1} + \gamma\ket{g_1}\right)\otimes\kb{\nu_0} + \rej \right)
\]
where, above, $\kb{\nu_0}$ is some state in Eve's ancilla used by her when the second sub-round is not used (without loss of power to Eve, this is a pure state) and $\rej$ is the state of the system in the case that Alice and Bob signal to discard this round (this state will later be projected out when we condition on this round's acceptance and so we do not bother to derive what it is).  We also define $P(\ket{z}) = \kb{z}$ to simplify the expressions.

Finally, if the server sends the message ``$0$'' parties run the second sub-round, flipping their actions to $\M$.  In this case, as discussed in our security model, Eve prepares, for sub-round 2, the same state as before, sending $\ket{\psi_0}$.  Alice and Bob both then $\M$ and discard if they see a photon.  Thus, focusing on the part of the state that will ultimately not be rejected (in particular, Alice and Bob do not observe the photon when it arrives, thus causing the state to collapse to $\ket{v,c_v}$ then returning to Eve), we find the final state for Alice, Bob, and Eve to be:
\begin{align*}
\kb{01}_{AB} \otimes (&\kb{1}_{cl}\otimes P\left(\alpha\ket{e_1} + \beta\ket{f_1} + \gamma\ket{g_1}\right)\otimes\nu_0\\
+& \kb{0}_{cl}\otimes P(\alpha\ket{e_0} + \beta\ket{f_0} + \gamma\ket{g_0})\otimes\left[\kb{0}_{cl}\kb{g_0} + \kb{1}_{cl}\kb{g_1}\right] + \rej),
\end{align*}
where the $AB$ registers are used to store Alice and Bob's classical raw key choice.  Note that $\ket{\nu_0}$ is a state in the Hilbert space used to model the classical message and Eve's ancilla in the second sub-round assuming that sub-round is not used.  Using similar techniques, one may trace the protocol for the other three cases of actions, leading to the following results (ignoring any ``reject'' states which, of course, appear  in all the cases below):
\begin{align*}
\kb{10}_{AB}\otimes(&\kb{1}_{cl}\kb{g_1}\otimes\kb{\nu_0}\\
+ &\kb{0}_{cl}\otimes\kb{g_0}\otimes [\kb{1}_{cl}\otimes P(\alpha\ket{e_1} + \beta\ket{f_1} + \gamma\ket{g_1}) + \kb{0}_{cl}\otimes P(\alpha\ket{e_0} + \beta\ket{f_0} + \gamma\ket{g_0})])\\
\kb{00}_{AB}\otimes(&\kb{1}_{cl}\otimes P \left( \alpha\ket{e_1} + \gamma\ket{g_1}\right) \otimes \kb{\nu_0}\\
+& \kb{0}_{cl}\otimes P(\alpha\ket{e_0} + \gamma\ket{g_0}) \otimes \left[\kb{0}_{cl}\otimes P(\beta\ket{f_0} + \gamma\ket{g_0}) + \kb{1}_{cl}\otimes P(\beta\ket{f_1} + \gamma\ket{g_0})\right]\\
\kb{11}_{AB}\otimes(&\kb{1}_{cl}\otimes P \left( \beta\ket{f_1} + \gamma\ket{g_1}\right) \otimes \kb{\nu_0}\\
+& \kb{0}_{cl}\otimes P(\beta\ket{f_0} + \gamma\ket{g_0}) \otimes \left[\kb{0}_{cl}\otimes P(\alpha\ket{e_0} + \gamma\ket{g_0}) + \kb{1}_{cl}\otimes P(\alpha\ket{e_1} + \gamma\ket{g_0})\right]
\end{align*}

To clean up the resulting density operator, we introduce the following notation:
\begin{align}
r_1&= \alpha \ket{e_1} + \beta \ket{f_1} +\gamma \ket{g_1}\label{eq:vectorstates}\\
r_0&= \alpha \ket{e_0} + \beta \ket{f_0} +\gamma \ket{g_0}\notag\\
s_1&= \beta \ket{f_1}+ \gamma \ket {g_1}\notag\\
s_0&= \beta \ket{f_0}+ \gamma \ket {g_0}\notag\\
t_1&= \alpha \ket{e_1}+ \gamma \ket {g_1}\notag\\
t_0 &= \alpha \ket{e_0}+ \gamma \ket {g_0}\notag
\end{align}
Using this, we derive the following density operator $\rho_{ABE}$ which models the entire joint state of the protocol conditioning on the round not being rejected (i.e., we now project out the ``reject'' states above and re-normalize):
\begin{align}
\rho_{ABE} = &\frac{1}{N}\kb{00}_{AB}\otimes( \kb{1, t_1, \nu_0} + \kb{0, t_0, 0, s_0} + \kb{0,t_0,1,s_1})\\
+&\frac{1}{N}\kb{11}_{AB}\otimes( \kb{1, s_1, \nu_0} + \kb{0, s_0, 0, t_0} + \kb{0,s_0,1,t_1})\notag\\
+&\frac{1}{N}\kb{01}_{AB}\otimes( \kb{1,r_1,\nu_0} + \kb{0,r_0,0,g_0} + \kb{0,r_0,1,g_1})\notag\\
+&\frac{1}{N}\kb{10}_{AB}\otimes( \kb{1,g_1,\nu_0} + \kb{0,g_0,0,r_0} + \kb{0,g_0,1,r_1})\notag
\end{align}
where $N$ is the normalization term:
\begin{align}
  N &= \bk{t_1} + \bk{t_0}\bk{s_0} + \bk{t_0}\bk{s_1}\label{eq:N}\\
  &+ \bk{s_1} + \bk{s_0}\bk{t_0} + \bk{s_0}\bk{t_1}\notag\\
  &+ \bk{r_1} + \bk{r_0}\bk{g_0} + \bk{r_0}\bk{g_1}\notag\\
  &+ \bk{g_1} + \bk{g_0}\bk{r_0} + \bk{g_0}\bk{r_1}.\notag
\end{align}

Our goal is to compute $H(A|E)_\rho$.  Applying Theorem \ref{thm:entropy} and simplifying the resulting expression yields:
\begin{align}
H(A|E)_\rho \ge& \frac{\bk{t_1} + \bk{s_1}}{N}\left[ H\left( \frac{\bk{t_1}}{\bk{t_1} + \bk{s_1}}\right) - H(\lambda_1)\right]\label{eq:full-entropy}\\
+& \frac{2\bk{t_0,s_0}}{N}\left[ 1 - H(\lambda_2)\right]\notag\\
+& \frac{\bk{t_0,s_1} + \bk{s_0,t_1}}{N}\left[ H\left( \frac{\bk{t_0,s_1}}{\bk{t_0,s_1} + \bk{s_0,t_1}}\right) - H(\lambda_3)\right]\notag\\
+& \frac{\bk{r_1} + \bk{g_1}}{N}\left[ H\left( \frac{\bk{r_1}}{\bk{r_1} + \bk{g_1}}\right) - H(\lambda_4)\right]\notag\\
+& \frac{2\bk{r_0,g_0}}{N}\left[ 1 - H(\lambda_5)\right]\notag\\
+& \frac{\bk{r_0,g_1} + \bk{g_0,r_1}}{N}\left[ H\left( \frac{\bk{r_0,g_1}}{\bk{r_0,g_1} + \bk{g_0,r_1}}\right) - H(\lambda_6)\right]\notag
\end{align}
where:
\begin{align*}
\lambda_1 &= \frac{1}{2}\left(1 + \frac{\sqrt{(\bk{t_1} - \bk{s_1})^2+4|\braket{t_1|s_1}|^2}}{(\bk{t_1} + \bk{s_1})}\right)\\
\lambda_2 &= \frac{1}{2}\left(1 + \frac{|\braket{t_0,s_0|s_0,t_0}|}{\bk{t_0,s_0}}\right)\\
\lambda_3 &= \frac{1}{2}\left(1 + \frac{\sqrt{(\bk{t_0,s_1} - \bk{s_0,t_1})^2+4|\braket{t_0,s_1|s_0,t_1}|^2}}{(\bk{t_0,s_1} + \bk{s_0,t_1})}\right)\\
\lambda_4 &= \frac{1}{2}\left(1 + \frac{\sqrt{(\bk{r_1} - \bk{g_1})^2+4|\braket{r_1|g_1}|^2}}{(\bk{r_1} + \bk{g_1})}\right)\\
\lambda_5 &= \frac{1}{2}\left(1 + \frac{|\braket{r_0,g_0|r_0,g_0}|}{\bk{r_0,g_0}}\right)\\
\lambda_6 &= \frac{1}{2}\left(1 + \frac{\sqrt{(\bk{r_0,g_1} - \bk{g_0,r_1})^2+4|\braket{r_0,g_1|g_0,r_1}|^2}}{(\bk{r_0,g_1} + \bk{g_0,r_1})}\right)\\
\end{align*}

We must now show how those inner products appearing in the above expression may be bounded through observable quantities (i.e., through the probabilities of certain observable events occurring).  This will allow us to calculate a lower-bound on the key-rate of our protocol based only on observable quantities.

\subsection{Parameter Estimation and Evaluation}

We can determine bounds on the needed inner products by considering various, observable, events such as the probability of the server sending the message ``$1$'' given that both parties choose $\R$ (this should be small for instance).  This can be done for arbitrary channels, however to actually evaluate and compare to prior work, we will also derive expressions for a symmetric depolarization attack, a common approach in QKD security proofs.  Note that our security proof above and below does not require this as an assumption - it is only done in order to evaluate the performance on a standard channel scenario.  Our steps below, however, may be followed for any observed channel.  We also consider symmetric attacks of this form in order to compare with prior work and protocols, the majority of which consider depolarization channels in their evaluation.  Under these evaluation conditions, we may parameterize the channel statistics as follows: $\phi$ will be the phase error of the channel; $p_l$ is the probability of loss in one direction (the server to users and the users to the server); and $p_d$ is the dark count rate of the server's detectors (note that if the server is adversarial, it may have perfect detectors, but try to ``hide'' its attack by simulating a suitable dark count rate).

We begin by considering $Pr(C = 1 \given A=B=\R) = P_{1|RR}$ which is the probability that, conditioning on both Alice and Bob choosing $\R$, the server sends the message $1$.  It is clear, from the above analysis, that this is $P_{1|RR} = \bk{r_1}$.  Under our symmetric attack scenario, we set this to $P_{1|RR} = \frac{p_lp_d}{2} + (1-p_l)\left(\frac{p_lp_d}{2} + (1-p_l)\phi\right)$.  Similarly, we find the following:
\begin{align*}
P_{1|RR} &= \bk{r_1} = \frac{p_lp_d}{2} + (1-p_l)\left(\frac{p_lp_d}{2} + (1-p_l)\phi\right)\\
P_{0|RR} &= \bk{r_0} = \frac{p_lp_d}{2} + (1-p_l)\left(\frac{p_lp_d}{2} + (1-p_l)(1-\phi)\right)\\\\
P_{1|MR} &= \bk{s_1} = \frac{p_lp_d}{2} + \frac{1-p_l}{2}\left(\frac{p_lp_d}{2} + \frac{1-p_l}{2}\right)\\
P_{0|MR} &= \bk{s_0} = \frac{p_lp_d}{2} + \frac{1-p_l}{2}\left(\frac{p_lp_d}{2} + \frac{1-p_l}{2}\right)\\\\
P_{1|RM} &= \bk{t_1} = \frac{p_lp_d}{2} + \frac{1-p_l}{2}\left(\frac{p_lp_d}{2} + \frac{1-p_l}{2}\right)\\
P_{0|RM} &= \bk{t_0} = \frac{p_lp_d}{2} + \frac{1-p_l}{2}\left(\frac{p_lp_d}{2} + \frac{1-p_l}{2}\right)\\\\
P_{1|MM} &= \bk{g_1} = \frac{p_lp_d}{2}\\
P_{0|MM} &= \bk{g_0} = \frac{p_lp_d}{2}
\end{align*}
Note that, in the above, we are defining $P_{i|RM}=P_{i|MR}$ to be the probability of the server sending message $i$ \emph{and} the measuring party not detecting the photon.  These let us easily compute $N$ using Equation \ref{eq:N} and the above.
%\begin{align*}
%  N &= 2p_lp_d + (1-p_l)\left(p_d^2 + \frac{1-p_l}{2}\right) + (1-p_l)^2\phi + 4\left(p_lp_d + \frac{1-p_l}{2}\left(p_d^2+\frac{1-p_l}{2}\right)\right)^2\\
%  &+ p_d\left(1 + 3(1-p_l)^2(1-\phi) + (1-p_l)^2\phi\right).
%\end{align*}

It is also clear that the values of $\alpha$, $\beta$, and $\gamma$ may be observed based on Alice and Bob's measurements.  Namely, $|\alpha|^2$ is the probability that, conditioning on both parties choosing $\M$, that Alice detects the photon.  Similar observations may be made for $|\beta|^2$, while $|\gamma|^2$ is the probability that neither party detects a photon.  Thus, these are:
\[
|\alpha|^2 = |\beta|^2 = \frac{1-p_l}{2}; |\gamma|^2 = p_l,
\]

%\begin{equation}
%\braket{r_1|g_1} = \alpha Re\braket{e_1|g_1} + \beta Re\braket{f_1|g_1} + \gamma Re\braket{g_1|g_1}.
%\end{equation}
%From the Cauchy-Schwarz inequality and the reverse triangle inequality, we have:
%\[
%|\braket{r_1|g_1}| \ge (\alpha+\beta)\sqrt{d} - \gamma d.
%\]
%Note that, above, we used the fact that $|\braket{e_1|g_1}| \le \sqrt{\bk{e_1}\bk{g_1}} \le \sqrt{\bk{g_1}}$ since $\bk{e_1} \le 1$ (similarly for $\braket{f_1|g_1}$).  A similar bound may be found for $Re\braket{r_0|g_0}$:
%\[-\alpha \sqrt{d}-\beta \sqrt {d}+\gamma \cdot d \leq Re\braket{r_0|g_0} \leq \alpha \sqrt{d}+\beta \sqrt {d}+\gamma \cdot d.\]
Next, we bound $|\braket{s_1|t_1}|$ and $|\braket{s_0|t_0}|$.  From Equation \ref{eq:vectorstates}, we have:
\begin{align*}
\braket{r_1|r_1}&=\alpha^2\braket{e_1|e_1}+\beta^2 \braket{f_1|f_1}+\gamma^2 \braket{g_1|g_1}+2\alpha\beta Re\braket{e_1|f_1}+ 2\beta\gamma Re\braket{f_1|g_1}+2\gamma\alpha Re\braket{g_1|e_1}\\
\end{align*}
Thus,
\begin{align}
\alpha\beta  Re\braket{e_1|f_1}&=\frac{1}{2}(\braket{r_1|r_1}-\alpha^2\braket{e_1|e_1}-\beta^2 \braket{f_1|f_1}-\gamma^2\braket{g_1|g_1}-2\beta\gamma Re\braket{f_1|g_1}-2\gamma\alpha Re\braket{g_1|e_1})\notag\\
&=\frac{1}{2}(\braket{r_1|r_1}-\alpha^2\braket{e_1|e_1}-\beta^2 \braket{f_1|f_1}-\gamma^2\braket{g_1|g_1})-\beta\gamma Re\braket{f_1|g_1}-\gamma\alpha Re\braket{g_1|e_1}\label{eq:abe1f1}
\end{align}
Now, we can write $Re\braket{s_1|t_1}$ as:
\[Re\braket{s_1|t_1}= \alpha \beta Re\braket{f_1|e_1}+\beta \gamma Re\braket{f_1|g_1}+\alpha\gamma Re\braket{g_1|e_1}+\gamma^2 \braket{g_1|g_1}\]
By substituting in Equation \ref{eq:abe1f1} and noting that $Re\braket{e_1|f_1} = Re\braket{f_1|e_1}$, we have:
\begin{align}
Re\braket{s_1|t_1} =&\frac{1}{2}(\braket{r_1|r_1}-\alpha^2\braket{e_1|e_1}-\beta^2 \braket{f_1|f_1}-\gamma^2\braket{g_1|g_1})\notag\\
&-\beta\gamma Re\braket{f_1|g_1}-\gamma\alpha Re\braket{g_1|e_1}+\beta \gamma Re\braket{f_1|g_1}+\alpha\gamma Re\braket{g_1|e_1}+\gamma^2 \braket{g_1|g_1}\notag\\
    &=\frac{1}{2} \braket{r_1|r_1}-\frac{1}{2}\alpha^2\braket{e_1|e_1}-\frac{1}{2}\beta^2 \braket{f_1|f_1}+\frac{1}{2}\gamma^2\braket{g_1|g_1}\label{eq:s1t1-first}
\end{align}
Next we may find an expression for $\alpha^2\braket{e_1|e_1}$ by looking at $\braket{t_1|t_1}$ (which is an observable quantity as discussed above, namely $P_{1|RM}$):
\[
\braket{t_1| t_1}= \alpha^2 \braket{e_1 |e_1} + \gamma^2 \braket {g_1 |g_1} + 2 \alpha\gamma Re\braket{e_1 |g_1}
\]
\[
\Longrightarrow \alpha^2 \braket{e_1 |e_1}= \braket{t_1| t_1}- \gamma^2 \braket {g_1| g_1} - 2 \alpha\gamma Re \braket{e_1| g_1}
\]

Of course $\braket{t_1 |t_1}$ is an observable probability for Alice and Bob and, later, we may use Cauchy-Schwarz to bound $|\braket{e_1|g_1}|$ thus allowing them to bound $\alpha^2\braket{e_1|e_1}$ used in the expansion of $\braket{s_1|t_1}$.  Similarly, we find:
\[
\braket{s_1| s_1}= \beta^2 \braket{f_1| f_1} + \gamma^2 \braket {g_1|g_1} + 2 \beta\gamma Re \braket{f_1|g_1}
\]
\[
\Longrightarrow \beta^2 \braket{f_1|f_1}= \braket{s_1|s_1}- \gamma^2 \braket {g_1|g_1} - 2 \beta\gamma Re  \braket{f_1|g_1}
\]

Combining this into Equation \ref{eq:s1t1-first} and using the (reverse) triangle inequality yields:
\begin{align}
  |\braket{s_1|t_1}| \ge |Re\braket{s_1|t_1}|&= \frac{1}{2}\left|\bk{t_1} + \bk{s_1} - \bk{r_1} - 3\bk{g_1}- 2\alpha\gamma Re\braket{e_1|g_1} - 2\beta\gamma Re\braket{f_1|g_1}\right|\notag\\
  &\ge \frac{\bk{t_1} + \bk{s_1} - \bk{r_1}}{2} - \frac{3}{2}\bk{g_1} - (\alpha\gamma + \beta\gamma)\sqrt{\bk{g_1}},
\end{align}
where, for the last inequality, we used the fact that $|\braket{e_1|g_1}| \le \sqrt{\bk{e_1}\bk{g_1}} \le \sqrt{\bk{g_1}}$. (Similarly for $\braket{f_1|g_1}$.)
Similarly we may bound:
\begin{align}
  |\braket{s_0|t_0}| &\ge \frac{1}{2}\left|\bk{t_1} + \bk{s_1} - \bk{r_1} - 3\bk{g_1} - 2\alpha\gamma Re\braket{e_1|g_1} - 2\beta\gamma Re\braket{f_1|g_1}\right|\\
  &\ge \frac{\bk{t_0} + \bk{s_0} - \bk{r_1}}{2} - \frac{3}{2}\bk{g_1} - (\alpha\gamma + \beta\gamma)\sqrt{\bk{g_1}}.
\end{align}

The only remaining inner-products we require are $\braket{r_0|g_0}$ and $\braket{g_1|r_1}$.  However, we were unable to find a non-trivial bound for these based only on observed statistics.  Our analysis shows that Eve can always set these to be orthogonal states without inducing additional noise.  Note that by making these orthogonal, Eve has maximal information gain from these particular states.  Thus, to work around this, we take advantage of the fact that Theorem \ref{thm:entropy} allows us to remove summation terms \emph{while still generating a lower-bound on the entropy}.  Therefore, we will instead use the following entropy bound, which can only be lower than the one in Equation \ref{eq:full-entropy} (thus this bound gives more advantage to the adversary):
\begin{align}
H(A|E)_\rho \ge& \frac{\bk{t_1} + \bk{s_1}}{N}\left[ H\left( \frac{\bk{t_1}}{\bk{t_1} + \bk{s_1}}\right) - H(\lambda_1)\right]\label{eq:partial-entropy}\\
+& \frac{2\bk{t_0,s_0}}{N}\left[ 1 - H(\lambda_2)\right]\notag\\
+& \frac{\bk{t_0,s_1} + \bk{s_0,t_1}}{N}\left[ H\left( \frac{\bk{t_0,s_1}}{\bk{t_0,s_1} + \bk{s_0,t_1}}\right) - H(\lambda_3)\right]\notag
\end{align}
Though we don't use it in our evaluation, we keep Equation \ref{eq:full-entropy} in this paper to aid future researchers.  If it is possible to derive a non-trivial bound for those inner products appearing in $\lambda_4$, $\lambda_5$, or $\lambda_6$, the key-rate bound can only improve.  We derive a lower bound here that may not be optimal - yet, despite this, we show improved performance over the original protocol (as we soon show).

\begin{figure}
    \centering
    \includegraphics[width=.7\textwidth]{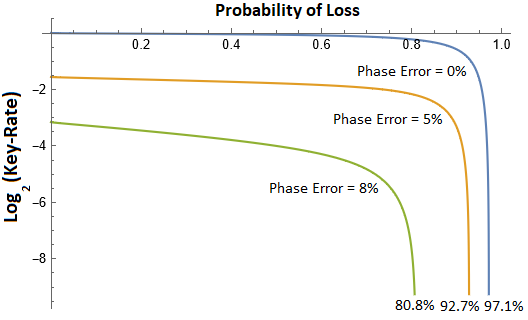}
    \caption{Evaluating our protocol's key-rate as the probability of loss increases for fixed phase error rate $\phi$.  Here we set $p_d = 10^{-6}$ (a typical value for detector dark counts).  Blue (top): $\phi = 0$; Yellow (middle): $\phi = 5\%$; Bottom (green): $8\%$. Note that this is assuming single qubits and lossy channels - if multi-photon attacks were analyzed the maximal supported probability of loss would be significantly lower; however potentially decoy state methods \cite{decoy1,decoy2,decoy3,decoy4} may be used to improve that though we leave that as interesting future work.}
    \label{fig:distance}
\end{figure}

\begin{figure}
    \centering
    \includegraphics[width=.7\textwidth]{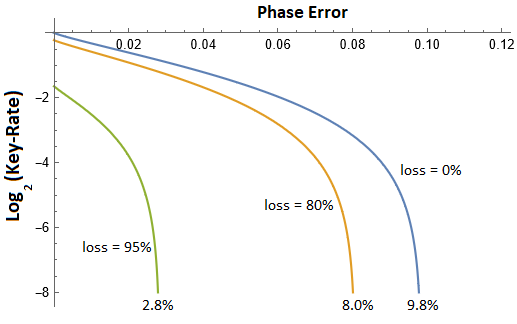}
    \caption{Evaluating our protocol's key-rate as the phase error rate $\phi$ increases, for fixed loss rate.  Blue (top): $p_l = 0$; Yellow (middle): $p_l = 0.8$; Green (bottom): $p_l = 0.95$. For all evaluations, we set $p_d = 10^{-6}$.}
    \label{fig:noise}
\end{figure}

This gives us everything we need to evaluate our key-rate bound.  In Figure \ref{fig:distance}, we show how our protocol behaves as the probability of loss increases while in Figure \ref{fig:noise}, we show how the protocol behaves when noise varies and see that the maximal phase noise $\phi$ allowed is $9.8\%$ when there is no loss (as the probability of loss increases, the maximal noise tolerance of course decreases as expected).  

\begin{figure}
    \centering
    \includegraphics[width=.7\textwidth]{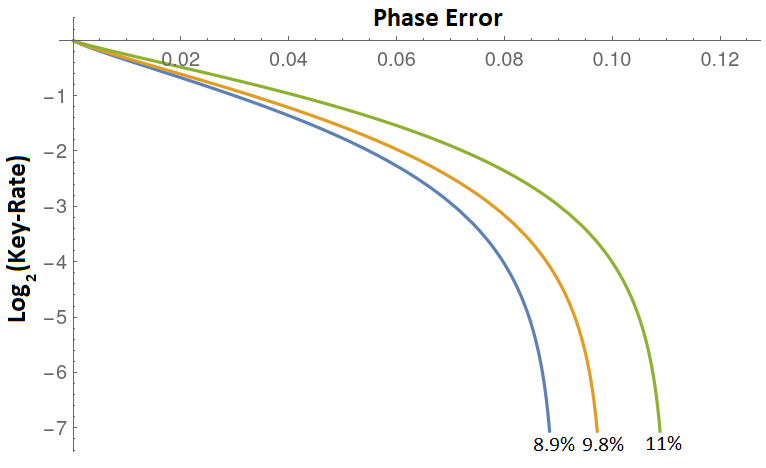}
    \caption{Comparing our protocol's key rate (Yellow, middle) with the original MZ-M-SQKD19 protocol in \cite{massa2019experimental} (Blue, bottom) with similar parameters; also comparing with BB84 (Green, top).  Here, we consider no loss and no dark counts, while we vary the phase error $\phi$.  We note that the extension we propose here has a higher noise tolerance and higher key-rate than the original MZ-M-SQKD19.}
    \label{fig:comp1}
\end{figure}

We also compare to the original MZ-M-SQKD19 protocol introduced in \cite{massa2019experimental} which our protocol extends.  For this comparison, we look at the ideal case of no loss and no dark counts for both.  This comparison is shown in Figure \ref{fig:comp1}.  We note that the key-rate is significantly improved for our new protocol for the same channel noise scenario.  We also compare to BB84's key rate of $1-2h(\phi)$ \cite{shor2000simple,QKD-renner-keyrate}.   Of course, BB84 outperforms as expected, however our extension does bring the key-rate closer to that of BB84 (in the ideal scenario which is all we consider here).

Next, we consider the effective key-rate which is defined to be the number of secret bits over the total number of signals sent. The previous graphs were the number of secret bits over the raw key size, a value that is higher than the effective rate as the effective rate takes into account rounds that were discarded and the fact that some rounds require two qubits as the second sub-round was invoked.  Let $Q$ be the number of photons sent (in the combined sub-round 1 and sub-round 2 for all used rounds) and let $M$ be the total number of rounds (where a round can consist of one or two sub-rounds; thus a round can contribute one or two photons to the total number of photons sent).  Note that with most QKD protocols, and in particular the original MZ-M-SQKD19 protocol, it holds that $Q=M$; but this is not the case for our protocol.  Finally, let $K$ be the size of the raw key.  We have computed, above, the ratio $\ell/K$ as $K \rightarrow \infty$.  We next derive $\ell/Q$ and for this, we must express $Q$ as a function of $K$.  Normally, these values may all be observed; however to evaluate this and compare we will again continue to assume our symmetric noise model.  This is not required, as mentioned before, it simply makes the algebra easier.  It is clear that $Q = M + M\cdot Pr(\text{Sub-Round $2$ is Used}) = M(1 + Pr(\text{$C$ sends ``0'' on Sub-Round 1}))$.  Let $p_0 =Pr(\text{$C$ sends ``0'' on Sub-Round 1})$; in our symmetric noise model, we find this to be:
\[
p_0 = \frac{1}{4}\left(2p_lp_d + (1-p_l)\left(p_lp_d + \frac{1-p_l}{2} + (1-p_l)(1-\phi)\right)\right)
\]
It is clear that $K = p_{acc}M$, where $p_{acc}$ is the probability of accepting any particular round (i.e., the probability that a round leads to a raw key bit generation).  This is easily seen to be $p_{acc} = \frac{1}{4}N$, where $N$ is given in Equation \ref{eq:N}.  Thus, combining everything, we have:
\[
K = \frac{p_{acc}Q}{1+p_0} = \frac{NQ}{4(1+p_0)} \Longrightarrow Q = \frac{4(1+p_0)K}{N},
\]
and so we find the effective key-rate $r' = \frac{\ell}{Q}$ to be:
\[
r' = \frac{\ell}{Q} = \frac{N\ell}{4(1+p_0)K} = \frac{N}{4(1+p_0)}r.
\]

\begin{figure}
    \centering
    \includegraphics[width=.7\textwidth]{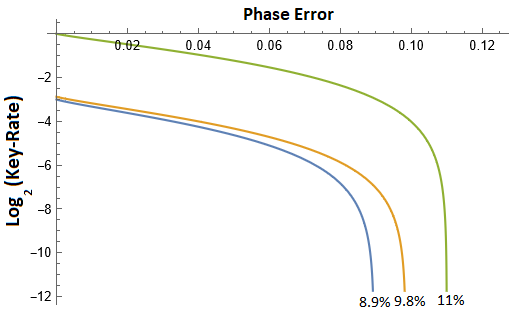}
    \caption{Comparing our protocol's \emph{effective} key rate (Yellow, middle) with the original MZ-M-SQKD19 protocol in \cite{massa2019experimental} (Blue, bottom) with similar parameters; also comparing with BB84 (Green, top).  Here we consider no loss and no dark counts while we vary the phase error $\phi$.  We note that even when considering the occasional need for two quantum signals per raw key bit (in the case a second sub-round is used), our extension is still more efficient and noise tolerant with similar parameters.}
    \label{fig:comp2}
\end{figure}

\begin{figure}
    \centering
    \includegraphics[width=.7\textwidth]{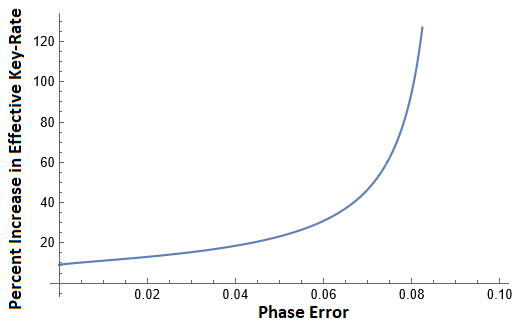}
    \caption{Showing the percent improvement in effective key rate of our extension compared to the original MZ-M-SQKD19 protocol under a noisy but lossless channel and no dark counts.  That is, we plot $\frac{r'_{new} - r'_{old}}{r'_{old}}$ for varying levels of phase noise $\phi$. Note that for $\phi > 8.9\%$, $r'_{old} = 0$ while our extension maintains a positive key-rate until $9.8\%$ thus the reason for the asymptote.}
    \label{fig:comp3}
\end{figure}

In Figure \ref{fig:comp2} we compare the effective key-rate of our extended protocol with the original MZ-M-SQKD19 protocol.  We note that, even when factoring in the need for an additional qubit, our extended version is still more efficient overall.  In Figure \ref{fig:comp3} we show the overall improvement between the two protocols.  We note that, as the noise increases, the percentage of increase in effective key-rate of our extension also increases.  Thus, our extension becomes highly useful the noisier the channel becomes.

\section{Closing Remarks}
%% comment on practical attacks

In this paper, we introduced an extension to the mediated SQKD protocol introduced in \cite{massa2019experimental}.  Our extension was designed to improve efficiency of the overall system by discarding fewer rounds.  Even though our extension occasionally requires the use of two signals per round, overall effective secret key rates are still improved even under noisy channels.  Interestingly, our extension also improves the noise tolerance of the protocol.

Many interesting open problems remain.  First, would be a full security analysis of general attacks - techniques from \cite{guskind2021mediated} in reducing mediated SQKD protocols to entanglement based versions may be useful, though those techniques do not immediately apply and some new insights are required.  Also of interest would be to extend the original protocol further in an effort to not waste the vacuum events.  One candidate protocol we may consider is to activate sub-round 2 if the server sends the message ``$0$'' \emph{or} ``$vac$'' on sub-round 1.  It is clear that this would be correct and lead to improved efficiency, especially on lossy channels.  We tried to analyze the security of this protocol, however the entropy expression contained over $18$ terms and the analysis became intractable, thus alternative methods may be required to rigorously prove security of this candidate extension.  Finally, we comment that this protocol contains a high level of asymmetry in error rates.  Referring to Table \ref{tbl:sub-round1} shows that the only way to get an error of Alice$=1$ and Bob$=0$ is through a dark-count event (which are typically small).  It would be interesting to see if this can be harnessed somehow to improve key-rates, perhaps even through a new classical process (e.g., a version of classical advantage distillation \cite{maurer1993secret,bae2007key,chau2002practical}) that takes into account this asymmetry.

\section*{Acknowledgments} WOK would like to acknowledge support from the National Science Foundation under grant number 1812070.  SM would like to acknowledge the support of National Science Foundation grant number CNS-1950600, which supported her during a summer REU at the University of Connecticut. 

%\bibliographystyle{unsrt}
%\bibliography{local}

\end{document}